\documentclass[floatfix,reprint,nofootinbib,amsmath,amssymb,epsfig,pre,floats,letterpaper,groupedaffiliation]{revtex4-1}

\usepackage{amsmath}
\usepackage{amssymb}
\usepackage{amsthm}
\usepackage{bm}
\usepackage{dcolumn}
\usepackage[english]{babel}
\usepackage{enumitem}
\usepackage{epstopdf}
\usepackage{graphicx}
\usepackage{inconsolata}
\usepackage{listings}
\usepackage{xcolor}
\usepackage{footmisc}
\usepackage{hyperref}

\newcommand{\beq}{\begin{equation}}
\newcommand{\eeq}{\end{equation}}

\newtheorem{theorem}{Theorem}
\newtheorem{lemma}[theorem]{Lemma}
\newtheorem{assumption}{Assumption}

\theoremstyle{definition}
\newtheorem{observation}{Observation}

\theoremstyle{definition}
\newtheorem{definition}{Definition}

\begin{document}

\title{Augur: a Decentralized Oracle and Prediction Market Platform (v2.0)}

\author{Jack Peterson}
\author{Joseph Krug}
\author{Micah Zoltu}
\author{Austin K. Williams}
\author{Stephanie Alexander}
\affiliation{Forecast Foundation}

\date{\today}

\begin{abstract}
Augur is a trustless, decentralized oracle and platform for prediction markets.  The outcomes of Augur's prediction markets are chosen by users that hold Augur's native Reputation token, who stake their tokens on the actual observed outcome and, in return, receive settlement fees from the markets.  Augur's incentive structure is designed to ensure that honest, accurate reporting of outcomes is always the most profitable option for Reputation token holders.  Token holders can post progressively-larger Reputation bonds to dispute proposed market outcomes.  If the size of these bonds reaches a certain threshold, Reputation splits into multiple versions, one for each possible outcome of the disputed market; token holders must then exchange their Reputation tokens for one of these versions.  Versions of Reputation which do not correspond to the real-world outcome will become worthless, as no one will participate in prediction markets unless they are confident that the markets will resolve correctly.  Therefore, token holders will select the only version of Reputation which they know will continue to have value: the version that corresponds to reality.
\end{abstract}

\maketitle

Augur is a trustless, decentralized oracle and prediction market platform.  In a prediction market, individuals can speculate on the outcomes of future events; those who forecast the outcome correctly win money, and those who forecast incorrectly lose money~\cite{Wolfers_2004, Surowiecki_2005, Hanson_2006}.  The price of a prediction market can serve as a precise and well-calibrated indicator of how likely an event is to occur~\cite{Pennock_2001, Manski_2004, Wolfers_2005, Goel_2010}.

Using Augur, people will have the ability to trade in prediction markets at very low cost.  The only significant expenses participants assume is compensation to market creators and to users that report on the outcomes of markets once the event has taken place.  The result is a prediction market where trust requirements, friction, and fees will be as low as competitive market forces can drive them.

Historically, prediction markets have been centralized.  The simplest way to aggregate trades in a prediction market is for a trustworthy entity to maintain a ledger; similarly, the simplest way to determine the outcome of an event and distribute payouts to traders is for an impartial, trusted judge to determine the outcomes of the markets.  However, centralized prediction markets have many risks and limitations: they do not allow global participation, they limit what types of markets can be created or traded, and they require traders to trust the market operator to not steal funds and to resolve markets correctly.

Augur aims to resolve markets in a fully decentralized way.  Decentralized, trustless networks, such as Bitcoin\cite{Nakamoto_2008} and Ethereum\cite{Buterin_2013}, eliminate the risk that self-interest will turn into corruption or theft.  The only role of the Augur developers is to publish smart contracts to the Ethereum network.  The Augur contracts are totally automated: the developers do not have the ability to spend funds that are held in escrow on-contract, do not control how markets resolve, do not approve or reject trades or other transactions on the network, cannot undo trades, cannot modify or cancel orders, etc.  The Augur \textit{oracle} allows information to be migrated from the real world to a blockchain without relying on a trusted intermediary.  Augur will be the world's first decentralized oracle.

\section{How Augur Works}
Augur markets follow a four-stage progression: \textit{creation}, \textit{trading}, \textit{reporting}, and \textit{settlement}.  Anyone can create a market based on any real-world event.  Trading begins immediately after market creation, and all users are free to trade on any market.  After the event on which the market is based has occurred, the outcome of the event is determined by Augur's oracle.  Once the outcome is determined, traders can close out their positions and collect their payouts.

Augur has a native token, Reputation (REP)\footnote{REP is an ERC777 token on the Ethereum network.}.  REP is needed by market creators and by reporters when they report on the outcome of markets created on the Augur platform.  Reporters report on a market by \textit{staking} their REP on one of the market's possible outcomes.  By doing this, the reporter declares that the outcome on which the stake was placed matches the real-world outcome of the market's underlying event.  The consensus of a market's reporters is considered the ``truth" for the purpose of determining the market's outcome.  If a reporter's report of a market's outcome does not match the consensus reached by the other reporters, Augur redistributes the REP staked on the non-consensus outcome by this reporter to the reporters that reported with the consensus.

By owning REP, and participating in the accurate reporting on the outcomes of events, token holders are entitled to a portion of the fees on the platform.  Each staked REP token entitles its holder to an equal portion of Augur's market fees.  The more REP a reporter owns, and reports correctly with, the more fees they will earn for their work in keeping the platform secure.

Although REP plays a central role in Augur's operations, it is not used to trade in Augur's markets. Traders are not required to participate in the reporting process, so they will never need to own or use REP.

\begin{figure*}
\includegraphics[width=0.8\textwidth]{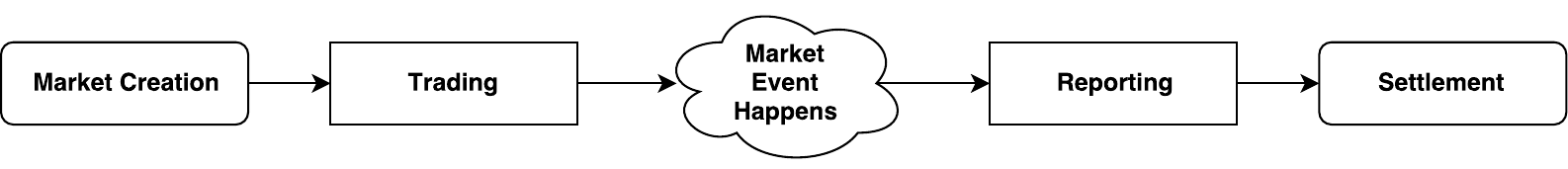}
\caption{Simplified outline of the lifetime of a prediction market.}
\label{fig:overview}
\end{figure*}

\subsection{Market Creation}

Augur allows anyone to create a market about any upcoming event. The \textit{market creator} sets the \textit{event end time} and chooses a \textit{designated reporter} to report the outcome of the event.  The designated reporter does not unilaterally decide the outcome of the market; the community always has an opportunity to dispute and correct the designated reporter's report.

Next, the market creator chooses a \textit{resolution source} that reporters should use to determine the outcome.  The resolution source may simply be ``common knowledge", or it may be a specific source, such as ``The United States Department of Energy", \texttt{bbc.com}, or the address of a particular API endpoint.\footnote{For example, if a market on ``The high temperature (in degrees Fahrenheit) on April 10, 2018 at the San Francisco International Airport, as reported by Weather Underground" specifies a resolution source of \texttt{https://www.wunderground.com/history/airport/KSFO/2018/4/10/ DailyHistory.html}, reporters would simply go to that URL and enter the high temperature displayed there as their report.}  They also set a \textit{creator fee}, which is the fee paid to the market creator by traders who settle with the market contract (see Section \ref{section:settlement} for details on fees).  Finally, the market creator posts two bonds: the \textit{validity bond}, and the \textit{creation bond}.

The validity bond is paid in DAI and is returned to the market creator if the market resolves to any outcome other than \textit{invalid}.\footnote{An \textit{invalid market} is a market determined to be invalid by reporters because none of the outcomes listed by the market creator is correct, or because the market wording is ambiguous or subjective; see Section \ref{section:ambiguous_or_subjective_markets} for discussion.}  The validity bond incentivizes market creators to create markets based on well-defined events with objective, unambiguous outcomes.  The size of the validity bond is set dynamically, based on the proportion of invalid outcomes in recent markets.\footnote{See Appendix \ref{section:bond_size_adjustment_details_validity_bonds} for details.}

The creation bond, paid in REP, is returned to the market creator if and only if the market's designated reporter actually reports during the first 24 hours after the market's \textit{event end time} and if the market ends up resolving to the same outcome that the designated reporter reported. If the designated reporter does not submit their report during the allotted 24 hours window, then the market creator forfeits the no-show bond and it is given to the \textit{first public reporter} who reports on the market (see Section \ref{section:open_reporting}).  This incentivizes the market creator to choose a reliable designated reporter who will report truthfully -- which should help markets resolve quickly.

In the event that the designated reporter fails to report, the creation bond is given to the first public reporter in the form of stake on their reported outcome, so that the first public reporter receives the creation bond if and only if they report correctly.  As with the validity bond, the creation bond is adjusted dynamically based on the proportion of designated reporters who failed to report on time during the previous dispute window and the proportion of markets that resolve to some outcome other than the one reported by the designated reporter.\footnote{See Appendix \ref{section:bond_size_adjustment_details_no-show_bonds} for details.}

The market creator creates the market and posts all required bonds via a single Ethereum transaction.  Once the transaction is confirmed, the market is live and trading begins.

\subsection{Trading}

Market participants forecast the outcomes of events by trading \textit{shares} of those market outcomes.  A \textit{complete set of shares} is a collection of shares that consists of one share of each possible valid outcome of the event~\cite{Clark_2014}.  Complete sets are created by Augur's on-contract matching engine as needed to complete trades\footnote{In practice, each share in a complete set is an ERC777 token on the Ethereum network.}.

For example, consider a market that has two possible outcomes, \texttt{A} and \texttt{B}.  Alice is willing to pay 0.7 DAI for a share of \texttt{A} and Bob is willing to pay 0.3 DAI for a share of \texttt{B}. First, Augur matches these orders and collects a total of 1 DAI from Alice and Bob.\footnote{The 1 DAI figure is used here for ease of discussion.  The actual cost of a complete set of shares is much smaller than this; see \texttt{docs.augur.net/\#number-of-ticks} for details.\label{footnote:complete_set_cost}}  Then Augur creates a complete set of shares, giving Alice the share of \texttt{A} and Bob the share of \texttt{B}.  This is how shares of outcomes come into existence.  Once the shares are created, they can be traded freely.

The Augur trading contracts maintain an order book for every market created on the platform.  Anybody can create a new order or fill an existing order at any time.  Orders are filled by an automated matching engine that exists within Augur's smart contracts.  Requests to buy or sell shares are fulfilled immediately if there is a matching order already on the order book.  It may be filled by buying shares from or selling shares to other participants, which, may involve issuing new complete sets or closing out existing complete sets.  Augur's matching engine always sequesters the minimum amount of shares and/or cash needed to cover the value at risk.  If there is no matching order, or the request can be only partially filled, the remainder is placed on the order book as a new order.

Orders are never executed at a worse price than the limit price set by the trader, but may be executed at a better price.  Unfilled and partially-filled orders can be removed from the order book by the order's creator at any time.  Fees are paid by traders only when complete sets of shares are sold; settlement fees are discussed in more detail in Section \ref{section:settlement}.

While most trading of shares is expected to happen before market settlement, shares can be traded any time after market creation.  All Augur assets -- including shares in market outcomes, participation tokens, shares in dispute bonds, and even ownership of the markets themselves -- are transferable at all times. In practice, all of these assets take the form of ERC777 tokens.

\subsection{Reporting}\label{section:reporting}

Once a market's underlying event occurs, the outcome must be determined in order for the market to finalize and begin settlement.  Outcomes are determined by Augur's oracle, which consists of profit-motivated reporters, who simply report the actual, real-world outcome of the event.  Anyone who owns REP may participate in the reporting and disputing of outcomes.  Reporters whose reports are consistent with objective reality are financially rewarded, while those whose reports are not consistent with objective reality are financially penalized (see Section \ref{section:rep_redistribution}).

\subsubsection{Dispute Windows}

Augur's reporting system runs on a cycle of consecutive 7-day long \textit{dispute windows}.  All fees collected by Augur during a given dispute window are added to the \textit{reporting fee pool} for that dispute window.  At the end of the dispute window, the reporting fee pool is paid out to REP holders who participated in the reporting process.  Reporters receive rewards in proportion to the amount of REP they staked during that dispute window.  Participation includes: staking during an initial report, disputing a tentative outcome, or purchasing \textit{participation tokens}.

\subsubsection{Participation Tokens}

During any dispute window, REP holders may purchase any number of participation tokens\footnote{Participation tokens are ERC777 tokens on the Ethereum network.} for one attorep\footnote{One \textit{attorep} is $10^{-18}$ REP.} each.  At the end of the dispute window, they may redeem their participation tokens for one attorep each, in addition to a proportional share of the dispute window's \textit{reporting fee pool}.  If there were no actions (\textit{e.g.}, submitting a report or disputing a report submitted by another user) needed of a reporter, the reporter may purchase participation tokens to indicate that they showed up for the dispute window.  Just like staked REP, participation tokens may be redeemed by their owners for a \textit{pro rata} portion of fees in this dispute window.

Participation tokens are primarily the means by which Augur pays fees to REP holders, but they may also serve as an additional incentive for REP holders to monitor the platform at least once per week. Even REP holders who do not want to participate in the reporting process may be incentivized to check-in with Augur once per 7-day dispute window in order to buy participation tokens and collect fees. This regular, active checking-in will ensure that they are familiar with how to use Augur, will be aware of forks when they occur, and thus should be more ready to participate in forks when they happen.

\subsubsection{Market State Progression}

Augur markets can be in seven different states after creation.  The potential states, or ``phases", of an Augur market are as follows:
\begin{itemize}
\item Pre-reporting
\item Designated Reporting
\item Open Reporting
\item Dispute Round
\item Waiting for Window
\item Fork
\item Finalized
\end{itemize}

\begin{figure*}
\includegraphics[width=0.8\textwidth]{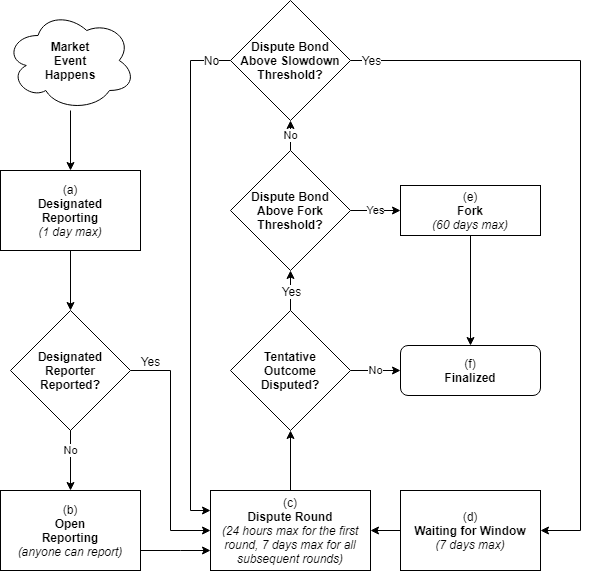}
\caption{Reporting flowchart.}
\label{fig:reporting}
\end{figure*}

The relationship between these states can be seen in Fig.~\ref{fig:reporting}.

\subsubsection{Pre-reporting}

The \textit{pre-reporting} or \textit{trading} phase (Fig.~\ref{fig:overview}) is the time period that begins after trading has begun in the market, but before the market's event has come to pass.  Generally, this is the most active trading period for any given Augur market.  Once the event end date has passed, the market enters the \textit{designated reporting} phase (Fig.~\ref{fig:reporting}a).

\subsubsection{Designated Reporting}

When creating a market, market creators are required to choose a designated reporter and post a creation bond.  During the designated reporting phase (Fig.~\ref{fig:reporting}a) the market's designated reporter has up to 24 hours to report on the outcome of the event.  If the designated reporter fails to report within the allotted 24 hours, the market creator forfeits the creation bond, and the market automatically enters the \textit{open reporting} phase (Fig.~\ref{fig:reporting}b).

If the designated reporter submits a report on time then the creation bond is placed as stake on the reported outcome, which will be forfeited if the market finalizes to any outcome other than the one they reported.\footnote{Forfeited stake is used to reward honest reporters and disputers; see Section \ref{section:rep_redistribution} for details.}  As soon as the designated reporter submits its report, the market enters the \textit{dispute round} phase (Fig.~\ref{fig:reporting}c), and the reported outcome becomes the market's \textit{tentative outcome}.

\subsubsection{Open Reporting}\label{section:open_reporting}

If the designated reporter fails to report within the allotted 24 hours, the market creator forfeits the creation bond, and the market immediately enters the \textit{open reporting} phase (Fig.~\ref{fig:reporting}b).  As soon as the market enters the open reporting phase, anyone can report the outcome of the market.  When the designated reporter fails to report, the first reporter who reports on the outcome of a market is called the market's \textit{first public reporter}.

The market's first public reporter receives the forfeited creation bond in the form of stake on their chosen outcome, so they may claim the no-show bond only if their reported outcome agrees with the market's final outcome.

The first public reporter does \textit{not} need to stake any of their own REP when reporting the outcome of the market.  In this way, any market whose designated reporter fails to report is expected to have its outcome reported by \textit{someone} very soon after entering the open reporting phase.

Once an \textit{initial report} has been received by the initial reporter (whether it was the designated reporter or first public reporter), the reported outcome becomes the market's tentative outcome, and the market enters the dispute round phase (Fig.~\ref{fig:reporting}c).

\subsubsection{Dispute Round}

The dispute round (Fig.~\ref{fig:reporting}c) is a phase during which any REP holder has the opportunity to dispute the market's \textit{tentative outcome}. A dispute round may last up to 7 days (with the exception of the very first dispute round, which may last up to 24 hours). At the beginning of a dispute round, a market's tentative outcome is the outcome that will become the market's final outcome if it is not successfully disputed by REP holders. A dispute consists of \textit{staking} REP (referred to as \textit{dispute stake} in this context) on an outcome \textit{other than} the market's current tentative outcome.  A dispute is \textit{successful} if the total amount of dispute stake on some outcome meets the \textit{dispute bond size} required for the current round.  The dispute bond size is computed as follows.

Let $A_n$ denote the total stake over all of this market's outcomes at the beginning of dispute round $n$.  Let $\omega$ be any market outcome \textit{other than} the market's tentative outcome at the beginning of this dispute round.  Let $S(\omega, n)$ denote the total amount of stake on outcome $\omega$ at the beginning of dispute round $n$.  Then the size of the \textit{dispute bond} needed to successfully dispute the current tentative outcome in favor of the new outcome $\omega$ during round $n$ is denoted $B(\omega, n)$ and is given by:
\beq \label{eq:bond_size}
B(\omega, n) = 2A_n - 3S(\omega, n)
\eeq

The bond sizes are chosen this way to ensure a fixed ROI for reporters who successfully dispute false outcomes (see Section \ref{section:leveraging_the_threat_of_a_fork}).

The dispute bonds need not be paid in their entirety by a single user.  The Augur platform allows participants to crowdsource dispute bonds.  Any user who sees an incorrect tentative outcome can dispute that outcome by staking REP on an outcome other than the tentative outcome. If any outcome (other than the tentative outcome) accumulates enough dispute stake to fill its dispute bond, the current tentative outcome will be successfully disputed.

In the case of a successful dispute, One of three things will happen: the market will either undergo another dispute round immediately, the market will wait until the next dispute window begins before undergoing another dispute round, or the market will enter the \textit{fork} state (Fig.~\ref{fig:reporting}e).  If the size of the filled dispute bond is greater than or equal to 2.5\% of all theoretical REP\footnote{All theoretical REP means the total theoretical supply of REP in the univese. In other words, Sum of the total amount of REP which exist in the universe and the total amount of REP which exist in the other universe and can be migrated to the universe.}, then the market will enter the fork state. If the size of the filled dispute bond is less than 2.5\% of all theoretical REP but greater than or equal to 0.02\% of all theoretical REP, then the newly chosen outcome becomes the market's new tentative outcome and the market enters the \textit{waiting for window} phase (Fig.~\ref{fig:reporting}d). If the size of the filled dispute bond is less than 0.02\% of all theoretical REP, then the newly chosen outcome becomes the market's new tentative outcome and the market immediately enters another dispute round.

All dispute stake is held in escrow during the dispute round. If a dispute bond is unsuccessful, then the dispute stake is returned to its owners at the end of the dispute round. If no dispute is successful during the dispute round, then the market enters the \textit{finalized} state (Fig.~\ref{fig:reporting}f), and its tentative outcome is accepted as its \textit{final outcome}.  A market's final outcome is the tentative outcome that passes through a dispute round without being successfully disputed, or is determined via a fork.  Augur's contracts treat final outcomes as \textit{truth} and pay out accordingly.

All unsuccessful dispute stake is returned to the original owners at the end of every dispute round.  All successful dispute stake is applied to the outcome it championed, and remains there until the market is finalized (or until a fork occurs in some other Augur market).  All dispute stake (whether successful or unsuccessful) will receive a portion of the \textit{reporting fee pool}\footnote{Any reporting fees and validity bonds collected during a dispute window get added to that dispute window's reporting fee pool.  At the end of the dispute window, the reporting fee pool is paid out to users in proportion to the amount of REP they staked during that dispute window.} from the current dispute window.

\subsubsection{Waiting for Window}

If a market's tentative outcome is disputed with a bond greater than or equal to 0.02\% of all theoretical REP but less than 2.5\% of all theoretical REP, then the market enters the waiting for window phase (Fig.~\ref{fig:reporting}d), before undergoing another dispute round. The purpose of this is simply to slow down the dispute process as the bonds get larger -- giving honest participants more time to crowdfund the larger dispute bond. This reduces the risk of a critical failure mode: one where the oracle resolves incorrectly because honest participants didn't have time to raise the funds needed to dispute a false tentative outcome.

During this phase, reporting for the market is on hold until end of the current dispute window.  Once the next dispute window begins, the market enters the \textit{dispute round} phase.

\subsubsection{Fork}\label{section:fork}

The fork state (Fig.~\ref{fig:reporting}e) is a special state that lasts up to 60 days.  Forking is the market resolution method of last resort; it is a very disruptive process and is intended to be a rare occurrence.  A fork is caused when there is a market with an outcome with a successfully-filled dispute bond of at least 2.5\% of all theoretical REP.  This market is referred to as the \textit{forking market}.

When a fork is initiated, a 60-day\footnote{Forking periods can be less than 60 days: a forking period ends when either 60 days have passed, or more than 50\% of all theoretical REP is migrated to some child universe. However even after the end of a forking period, REP in the parent universe can be migrated to the child universe if it is within 60 day of fork initiation.} \textit{forking period} begins.  Disputing for all other non-finalized markets is put on hold until the end of this forking period.  The forking period is much longer than the usual dispute window because the platform needs to provide ample time for REP holders and service providers (such as wallets and exchanges) to prepare.  A fork's final outcome cannot be disputed.

Every Augur market and all REP tokens exist in some \textit{universe}.  REP tokens can be used to report on outcomes (and thus earn fees) \textit{only} for markets that exist in the same universe as the REP tokens.  When Augur first launches, all markets and all REP will exist together in the \textit{genesis universe}.

When a market forks, new universes are created.  Forking creates a new \textit{child universe} for each possible outcome of the forking market (including \texttt{Invalid}).  For example, a ``Yes/No" market has 3 possible outcomes: \texttt{Yes}, \texttt{No}, and \texttt{Invalid}.  Thus, a ``Yes/No" forking market will create three new child universes: universe \texttt{Yes}, universe \texttt{No}, and universe \texttt{Invalid}.  Initially, these newly created universes are empty: they contain no markets or REP tokens.

When a fork is initiated, the \textit{parent universe} becomes permanently \textit{locked}.  In a locked universe, no new markets may be created.  Users may continue trading shares in markets in locked universes, and markets in a locked universe may still receive their initial reports. However, no reporting rewards are paid out there, and markets in locked universes cannot be finalized.  In order for markets or REP tokens in the locked universe to be useful, they must first be migrated to a child universe.

During the forking period, holders of REP tokens in the parent universe may migrate their tokens to a child universe of their choice.  This choice should be considered carefully, because migration is one-way; it cannot be reversed.  Tokens cannot be sent from one sibling universe to another.  \textit{Migration is a permanent commitment of REP tokens to a particular outcome of the forking market.}  REP tokens that migrate to different child universes ought to be considered entirely separate tokens, and service providers like wallets and exchanges ought to list them as such.

Any REP tokens that have not been migrated out of the parent universe 60 days after the fork started will be permanently locked in the parent universe. Such tokens are expected to lose all value, so it is of paramount importance that REP holders migrate their tokens anytime a fork happens.

When a fork is initiated, all REP staked on all non-forking markets is \textit{unstaked} so that it is free to be migrated to a child universe during the forking period.\footnote{The only exception is the REP staked by the initial reporter when they made the initial report.  That REP remains staked on the initial reported outcome and is automatically migrated to the child universe that wins the fork. This happens for technical reasons unrelated to the mechanism design.}

Whichever child universe receives the most migrated REP by the end of the forking period becomes the \textit{winning universe}, and its corresponding outcome becomes the final outcome of the forking market.  Un-finalized markets in the parent universe may be migrated only to the winning universe and, if they have received an initial report, are reset back to the waiting for window phase.

\paragraph*{Reporters that have staked REP on one of the forking market's outcomes cannot change their position during a fork.}  REP that was staked on a forking market's outcome in the parent universe can be migrated only to the child universe that corresponds to that outcome. For example, if a reporter helped fulfill a successful dispute bond in favor of outcome \texttt{A} during some dispute round, then the REP they have staked on outcome \texttt{A} can only be migrated to universe \texttt{A} during a fork.

\paragraph*{Sibling universes are entirely disjoint.}  REP tokens that exist in one universe cannot be used to report on events or earn rewards from markets in another universe.  Since users presumably will not want to create or trade on markets in a universe whose oracle is untrustworthy, REP that exists in a universe that does not correspond to objective reality is unlikely to earn its owner any fees, and therefore should not hold any significant market value.  Therefore, REP tokens migrated to a universe which does not correspond to objective reality should hold no market value, regardless of whether or not the objectively false universe ends up being the winning universe after a fork.  This has important security consequences, which we discuss in Section \ref{section:incentives_and_security}.

\subsubsection{Finalized}

A market enters the finalized state (Fig.~\ref{fig:reporting}f) if it passes through a 7-day dispute round without having its tentative outcome successfully disputed, or after completion of a fork.  The outcome of a fork cannot be disputed and is always considered final at the end of the forking period.  Once a market is finalized, traders can settle their positions directly with the market.  When a market enters the finalized state, we refer to its chosen outcome as the \textit{final outcome}.

\subsection{Market Settlement}\label{section:settlement}

A trader can close their position in one of two ways: by selling the shares they hold to another trader in exchange for currency, or by settling their shares with the market.  Recall that every share comes into existence as part of a complete set when a total of 1 DAI has been escrowed with Augur.\footref{footnote:complete_set_cost}  To get that 1 DAI out of escrow, traders must give Augur either a complete set or, if the market has finalized, a share of the winning outcome.  When this exchange happens we say traders are \textit{settling with the market contract}.

For example, consider a non-finalized market with possible outcomes \texttt{A} and \texttt{B}.  Suppose Alice has a share of outcome \texttt{A} that she wants to sell for 0.7 DAI and Bob has a share of outcome \texttt{B} that he wants to sell for 0.3 DAI. First, Augur matches these orders and collects the \texttt{A} and \texttt{B} shares from the participants. Then Augur gives 0.7 DAI (minus fees) to Alice and 0.3 DAI (minus fees) to Bob.

As a second example, consider a finalized market whose winning outcome is \texttt{A}.  Alice has a share of \texttt{A} and wants to cash it in.  She sends her share of \texttt{A} to Augur and in return receives 1 DAI (minus fees).

\subsubsection{Settlement Fees}

The only time Augur levies fees is when market participants are settling with the market contract.  Augur levies two fees during settlement: the creator fee, and the reporting fee. Both of these fees are proportional to the amount being paid out.  So, in the pre-finalized settlement example above, where Alice receives 0.7 DAI and Bob receives 0.3 DAI, Alice would pay 70\% of the fees while Bob would pay 30\%.

The creator fee is set by the market creator during market creation, and is paid to the market creator upon settlement.  The reporting fee is set dynamically (see Section \ref{section:market_cap_nudges}) and is paid to reporters who participate in the reporting process.

\subsubsection{Reputation Redistribution}\label{section:rep_redistribution}

If a market finalizes without initiating a fork, all REP staked on any outcome other than the market's final outcome is forfeited. Twenty percent of the forfeited stake is burned, and the remainder is distributed to the users who staked on the market's final outcome in proportion to the amount of REP they staked.  The dispute bond sizes and the amount burned are chosen such that anyone who successfully disputes an outcome in favor of the market's final outcome is rewarded with a 40\% ROI on their dispute stake.\footnote{See Theorem \ref{th:roi_guarantee} in Appendix \ref{section:finalization_time}.}  This is a strong incentive for reporters to dispute false tentative outcomes.

\section{Incentives and Security}\label{section:incentives_and_security}

There is a strong relationship between the market cap of REP and the trustworthiness of Augur's forking protocol.  If the market cap of REP is large enough\footnote{See Section \ref{section:integrity_forking_protocol} for details.}, and attackers are economically rational, then the outcome that wins the fork should correspond to objective reality.  In fact, it would be possible for Augur to function properly without using designated reporters and dispute rounds.  Using \textit{only} the forking process, the oracle would report truthfully.

However, forks are disruptive and time consuming.  A fork takes up to 60 days to resolve a single market, and can resolve only one market at a time.  During the 60 days in which the forking market is being resolved, all other non-finalized markets are put on hold.\footnote{Traders can continue trading on those markets, but those markets cannot finalize until after the forking period.}  Service providers have to update, and REP holders have to migrate their REP to one of the new child universes.  Therefore, forks should be used only when they are absolutely necessary.  Forking is the nuclear option.

Fortunately, once it has been established that forks can be trusted to determine truth, incentives can be used to encourage participants to behave honestly without having to actually initiate a fork.  \textit{It is the credible threat of a fork, and the belief that the fork will resolve correctly, that are the cornerstones of Augur's incentive system.}

Next, we discuss the conditions under which the forking system can be trusted to determine truth.  We then discuss the incentive system and how it encourages quick and correct resolution of all markets.

\subsection{Integrity of the Forking Protocol}\label{section:integrity_forking_protocol}

Here we discuss the reliability of the forking process and the conditions under which it can be trusted.  For ease of discussion, when referring to forks, we will refer to the child universe that corresponds to objective reality as the \texttt{True} universe, and any other child universe as a \texttt{False} universe.  We will refer to the child universe which receives the most REP migration during the forking period as the winning universe and all other child universes as losing universes.

Naturally, we always want the \texttt{True} universe to be the winning universe, and the \texttt{False} universes to be the losing universes.  We say that the forking protocol has been successfully attacked whenever a \texttt{False} universe ends up being the winning universe of a fork -- thus resulting in the forking market (and, potentially, all non-finalized markets) being paid out incorrectly.

Our approach to securing the oracle is to arrange matters such that the maximum benefit to a successful attacker is less than the minimum cost of performing the attack.  We formalize this below.

\subsubsection{Maximum Benefit to an Attacker}

An attacker who successfully attacks the oracle would cause all non-finalized Augur markets to migrate to a \texttt{False} universe.  If the attacker controls the majority of REP in the \texttt{False} universe, the attacker can then force all non-finalized markets to resolve however she wants.  In the most extreme case, she would also be able to capture all funds escrowed in all of those markets.\footnote{This would require the attacker to capture \textit{all} shares of some given outcome, and then force the market to finalize to that outcome.}

\begin{definition}
We define, and denote by $I_a$, Augur's \textit{native open interest} as the value of the sum of all funds escrowed in unfinalized Augur markets.\footnote{This includes external markets that pay reporting fees to Augur.}
\end{definition}

\begin{definition}
We define a \textit{parasitic market} as any market that does not pay reporting fees to Augur, but does resolve in accordance with the resolution of a native Augur market.
\end{definition}

\begin{definition}
We define, and denote by $I_p$, the \textit{parasitic open interest} as the value of the sum of all funds escrowed in all parasitic markets that resolve in accordance to non-finalized, native Augur markets.
\end{definition}

In the most extreme case, an attacker would also be able to capture all funds in all parasitic markets which resolve in accordance to non-finalized, native Augur markets.

\begin{observation}
The maximum (gross) benefit to an attacker who successfully attacks the oracle is $I_a + I_p$.
\end{observation}

\subsubsection{Parasitic Open Interest is Unknowable}

Augur can accurately and efficiently measure $I_a$.  However, $I_p$ cannot be known in general, as there may exist arbitrarily many offline parasitic markets, each with arbitrarily large open interest.  Since the maximum possible benefit to an attacker includes the unknowable quantity $I_p$, one can never be objectively certain that the oracle is secure against economically rational attackers.\footnote{This is true of all public oracles, even centralized ones.}

However, if we are willing to assert that $I_p$ is reasonably bounded in practice, then we can define conditions under which we may assert that the oracle is secure.

\subsubsection{Minimum Cost of a Successful Attack}

Next, consider the cost of attacking the oracle.  Let $P$ denote the price of REP. Let $\epsilon$ denote one attorep\footnote{One \textit{attorep} is $10^{-18}$ REP.}.  Let $M$ denote the total amount of REP in existence (the ``money supply" of REP).  Let $S$ denote the proportion of $M$ that will be migrated to the True universe during the forking period of a fork.

Thus the product $SM$ represents the absolute amount of REP migrated to the True universe during the forking period of a fork, and the product $PM$ is the market cap of REP.

Let $P_f$ denote the price of REP migrated to a \texttt{False} universe of the attacker's choosing.  Note that if $P \leq P_f$ then the oracle would not be secure against economically rational attackers, because it would be at least as profitable to migrate REP to the False universe as it would be to not migrate at all.

\subsubsection{Integrity}

\begin{assumption}
Reporters that are not attackers will never migrate REP to a \texttt{False} universe during a fork.\footnote{There may be cases where some non-malicious reporters do migrate REP to a \texttt{False} universe accidentally or carelessly.  However, such behavior is, in practice, indistinguishable from collaborating with an attacker.}
\end{assumption}

By design, a successful attack on the oracle requires more REP to be migrated to some \texttt{False} universe than to the \texttt{True} universe during the forking period of a fork.  By assumption, only the attacker will migrate REP to a \texttt{False} universe.  The amount of REP migrated to the \texttt{True} universe during the forking period is denoted by $SM$.  Thus, for an attacker to be successful, they must migrate at least $SM + \epsilon$ REP.  For simplicity, we will ignore the negligible $\epsilon$, and say that a successful attack requires migrating at least $SM$ REP, which has a value of $SMP$ before the migration, to some \texttt{False} universe.

If an attacker migrates $SM$ REP during the forking period of a fork, they will receive $SM$ REP on the child universe to which they migrate. If the attacker migrates to a \texttt{False} universe then the value of those coins becomes $SMP_f$.  Thus the minimum cost to the attacker is $(P - P_f)SM$.

\begin{observation}
The minimum amount of REP a successful attacker must migrate to a \texttt{False} universe during a fork is $SM$, which costs the attacker $(P - P_f)SM$.
\end{observation}

Note that if $S > \frac{1}{2}$ then an attack is \textit{impossible} because there does not exist enough REP outside of the \texttt{True} universe for any \texttt{False} universe to become the winning universe.

Pitted against economically rational attackers, the oracle will resolve to outcomes that correspond to objective reality if the maximum benefit to an attacker is less than the minimum cost of attack. By observations 1 \& 2 we can see that this occurs whenever $S > \frac{1}{2}$ or $I_a + I_p < (P - P_f)SM$.  This gives us our formal definition of integrity.

\begin{definition}\label{ob:integrity_property}
(Integrity Property) The forking protocol has \textit{integrity} whenever $S > \frac{1}{2}$ or whenever $I_a + I_p < (P - P_f)SM$.
\end{definition}

The above inequality can be solved for $PM$ to see the relationship between forking protocol integrity and the market cap of REP.

\begin{theorem}\label{th:market_cap_security_theorem}
(Market Cap Security Theorem) The forking protocol has integrity if and only if:
\begin{enumerate}
\item $S > \frac{1}{2}$, or
\item $P_f < P$ and the market cap of REP is greater than $\frac{(I_a + I_p)P}{(P - P_f)S}$.
\end{enumerate}
\end{theorem}

\begin{proof}
Suppose the forking protocol has integrity.  Then, by definition, $S > \frac{1}{2}$ or $I_a + I_p < (P - P_f)SM$. Suppose $I_a + I_p < (P - P_f)SM$. Since $I_a + I_p \geq 0$ and $SM > 0$, we know that $P_f < P$.  Then, solving $I_a + I_p < (P - P_f)SM$ for $PM$, we see that $\frac{(I_a + I_p)P}{(P - P_f)S} < PM$.  Thus the first direction is proved.

Now suppose that $S > \frac{1}{2}$, or that $P_f < P$ and $\frac{(I_a + I_p)P}{(P - P_f)S} < PM$. If $S > \frac{1}{2}$, then the forking protocol has integrity by definition.  If $P_f < P$ and $\frac{(I_a + I_p)P}{(P - P_f)S} < PM$, then, solving the inequality for $I_a + I_p$, we see that $I_a + I_p < (P-P_f) SM$, and the forking protocol has integrity.
\end{proof}

\subsection{Our Assumptions and Their Consequences}

We believe traders will not want to trade on Augur in a universe where reporters have lied.  We also believe that market creators will not pay to create Augur markets in a universe where there are no traders.  In a universe without markets or trading, REP does not provide any dividends to those holding it.  Therefore, we believe REP sent to a \texttt{False} universe will hold no non-negligible market value and we model this by letting $P_f = 0$.

An attacker should never migrate more than 50\% of all theoretical REP (plus on attorep) to a false universe\footnote{Doing so unnecessarily increases the cost of attack.}. We believe that approximately all theoretical REP will be migrated out of the parent universe during a forking period, because failure to do so is expected to result in loss of token value for no benefit. Therefore, we think it is reasonable to expect at least 50\% of REP (minus one attorep) to be migrated to the True outcome during the forking period of a fork, and we model this by letting $S \geq \frac{1}{2}$.  We are also willing to accommodate parasitic open interest as large as 50\% of the native open interest, and so we let $I_a \geq 2 I_p$.

Under these assumptions, Theorem \ref{th:market_cap_security_theorem} tells us that the forking protocol has integrity whenever the market cap of REP is at least 3 times the native open interest.

\subsection{Market Cap Nudges}\label{section:market_cap_nudges}

Augur gets information about the price of REP via a collection of trusted third-party price oracles\footnote{In the future, Augur may be able to learn the price of REP without having to trust third parties. This is an active area of research.}.  This gives Augur the ability to compute the current market cap of REP.  Augur can also measure the current native open interest, and can thus determine what market cap ought to be targeted in order to meet Augur's integrity requirements.

Every universe begins with a default reporting fee of 1\%. If the current market cap is below the target, then reporting fees are automatically increased (but will never be higher than 33.3\%), putting upward pressure on the price of REP and/or downward pressure on new native open interest.  If the current market cap is above the target, then reporting fees are automatically decreased (but will never be lower than 0.01\%) so that traders are not paying more than needed to keep the system secure.

The reporting fees are determined as follows.  Let $r$ be the reporting fee from the previous window, let $t$ be the target market cap, and let $c$ be the current market cap.  Then the reporting fee for the current dispute window is given by $\max\left\{ \min\left\{\frac{t}{c}r, \frac{333}{1000}\right\} , \frac{1}{10,000}\right\}$.

\subsection{Leveraging the Threat of a Fork}\label{section:leveraging_the_threat_of_a_fork}

As mentioned above, forks are a disruptive and slow way for markets to reach finalization.  Rather than using the forking process to resolve every market, Augur leverages the \textit{threat} of a fork to resolve markets efficiently~\cite{Williams_2019}.

Recall that any stake successfully disputing an outcome in favor of the market's final outcome will receive a 40\% ROI on their dispute stake.\footnote{Measured in REP that exists in a universe that corresponds to the market's final outcome; see Theorem \ref{th:roi_guarantee} in Appendix \ref{section:finalization_time}.}  In the event of a fork, any REP staked on any of the market's false outcomes should lose all economic value, while any REP staked on the market's true outcome is rewarded with 40\% more REP in the child universe that corresponds to the market's true outcome (regardless of the outcome of the fork).  Therefore, if pushed to a fork, REP holders who dispute false outcomes in favor of true outcomes will always come out ahead, while REP holders who staked on false outcomes will see their REP lose all economic value.

We believe this situation is sufficient to guarantee that all false tentative outcomes will be successfully disputed.

\section{Potential Issues \& Risks}

\subsection{Parasitic Markets}

Recall that a parasitic market is any market that does not pay reporting fees to Augur, but does resolve in accordance with the resolution of a native Augur market. Because parasitic markets do not have any reporters to pay, they can offer the same service as Augur with lower fees. This can have serious consequences for the integrity of Augur's forking protocol.

In particular, if parasitic markets attract trading interest away from Augur, then Augur's reporters will receive less in reporting fees. This would put downward pressure on the market cap of REP. If the market cap of REP falls too low, the integrity of the forking protocol is put in jeopardy (Theorem \ref{th:market_cap_security_theorem}). As a result, parasitic markets have the potential to threaten the long term viability of Augur, and should be vehemently opposed.

The oracle parasite problem has not been solved -- and may be provably unsolvable -- even for centralized systems. That said, at the time of this writing, we have observed no significant parasitic interest leveraging Augur's oracle. 

\subsection{Volatility of Open Interest}

Large, sudden, unexpected, and short-lived increases in open interest -- like those that may be seen during a popular sporting event -- result in rapid increases in the market cap requirement for forking protocol integrity (Theorem \ref{th:market_cap_security_theorem}).  When the market cap requirement exceeds the market cap, there is a risk of economically rational attackers causing a fork to resolve incorrectly.  While Augur does attempt to nudge the market cap and/or the open interest back into a safe ratio during such situations (see Section \ref{section:market_cap_nudges}), these nudges are reactionary and are adjusted only once per 7-day dispute window.

It is worth noting, however, that speculators who witness the sudden increase in open interest may buy REP in anticipation of the reactionary market cap nudge, thus driving the market cap of REP up, perhaps to a point where the integrity of the forking protocol is no longer threatened.  So the length of time during which the oracle is vulnerable may not be long enough for an attacker to successfully exploit the vulnerability.

Additionally, we have increased the security multiplier from the theoretical minimum of 3\footnote{The theoretical minimum for the security multiplier is 2 when not accounting for any parasitic interest. When account for parasitic interest of up to 50\% of the native open interest, the theoretical minimum for the security multiplier is 3.}, to a more comfortable value of 5. This means that the fee adjustment algorithm targets a market cap of 5 times the native open interest, rather than 3 times. This should help us absorb large (up to a 66.6\% increase), sudden, unexpected, and short-lived increases in open interest without threatening oracle integrity.

\subsection{Inconsistent or Malicious Resolution Sources}

During market creation, market creators chose a resolution source that reporters should use to determine the outcome of the event in question.  If the market creator chooses an inconsistent or malicious resolution source, honest reporters may lose money.

For example, suppose the market in question has outcomes \texttt{A} and \texttt{B}, and the market creator, Serena, has chosen her own website, \texttt{attacker.com}, as the resolution source.  After the market's event end time, Serena -- who is also the designated reporter for the market -- reports outcome \texttt{A}, and updates \texttt{attacker.com} to indicate that outcome \texttt{B} is the correct outcome.  Honest reporters who check \texttt{attacker.com} will see that the initial report is incorrect and, during the first dispute round, should successfully dispute the tentative outcome in favor of outcome \texttt{B}.  Serena would update \texttt{attacker.com} to indicate that outcome \texttt{A} is the correct outcome, and the market would then enter its second dispute round.  Again, reporters who check \texttt{attacker.com} will see that the tentative outcome (outcome \texttt{B}) is incorrect, and may successfully dispute it.  Serena can repeat this behavior until the market resolves.  No matter how the market resolves, some honest reporters will lose money.

Several variations of this attack exist.  Simply ignoring markets with dubious resolution sources is not sufficient, for in the event that such a market causes a fork, all REP holders will have to choose a child universe to which to migrate their REP.  Reporters should remain vigilant against markets with dubious resolution sources.  Such markets should be publicly identified so reporters can coordinate to make sure such markets finalize as invalid.

\subsection{Self-Referential Oracle Queries}

Markets that trade on the future behavior of Augur's oracle may have undesirable effects on the behavior of the oracle itself~\cite{Othman_2010}.  For example, consider a market that trades on the question, ``Will any designated reporter fail to submit a report during their three-day forking period before December 31, 2018?"  Bets placed on the \texttt{No} outcome of this market may act as a perverse incentive for designated reporters to intentionally fail to report.  If a designated reporter can buy up enough \texttt{Yes} shares at a low enough price to compensate for the loss of the creation bond, they may intentionally fail to report.

If the market cap of REP is large enough (Theorem \ref{th:market_cap_security_theorem}) then these self-referential oracle queries will not threaten the integrity of the forking protocol.  However, they may negatively affect the performance of Augur by causing delays in market finalizations.  While markets would still finalize correctly, this sort of behavior is disruptive and undesirable.

\subsection{Uncertain Fork Participation}\label{subsection:uncertain_fork_participation}

We cannot know in advance how much REP will be migrated to the \texttt{True} universe during the forking period of a fork, thus we cannot know in advance whether the market cap is large enough for the oracle to have integrity (Theorem \ref{th:market_cap_security_theorem}).  Our belief in the integrity of the forking protocol can be no stronger than our belief in our assumption that users will behave rationally and in their own economic self-interest. We assume that at least 50\% (minus one attorep) of all theoretical REP will migrate to the \texttt{True} child universe during the forking period of a fork because that is consistent with self-interested, rational decision making on the behalf of the participants. However, we cannot guarantee this will happen.

\subsection{Responsibility During a Fork}
Augur forks differ from blockchain forks in one important respect: after a blockchain fork, a user who owned a coin on the parent chain will now own a coin on both forks.  Ignoring replay attacks, blockchain forks pose little risk to users.  During an Augur fork, however, a user who owns a REP token in the parent universe can migrate that coin to only one of the child universes.  If the user migrates their token to any universe other than the consensus universe, their token may lose all value.  Thus migrating REP during the forking period of a fork, before it is clear which child universe has achieved consensus, exposes the user to risk.

This risk is an inherent part of REP ownership. Abstaining from migration during a fork will result in near-certain loss of value for the REP holder. This is a necessary design decision, as the integrity of the oracle relies upon REP holders reporting truthfully during the fork. REP holders cannot be absolved of this responsibility without greatly increasing the cost of system security.

\subsection{Ambiguous or Subjective Markets}\label{section:ambiguous_or_subjective_markets}

Only events that have objectively knowable outcomes are suitable for use in Augur markets.  If reporters believe that a market is not suitable for resolution by the platform -- for example, because it is ambiguous, subjective, or the outcome is not known by the event end date -- they should report the market as \texttt{Invalid}.  If a market resolves as \texttt{Invalid}, traders are paid out at equal values for all possible outcomes; for scalar markets, traders are paid out halfway between the market's minimum price and maximum price.

It is possible to imagine markets where some reporters are certain that the outcome is \texttt{A} and others are certain that the outcome is \texttt{B}.  For example, in 2006, TradeSports allowed its users to speculate on whether North Korea would fire a ballistic missile that would land outside of its airspace before the end of July 2006.  On July 5, 2006, North Korea successfully fired a ballistic missile that landed outside of its airspace, and the event was widely reported by the world media and confirmed by many U.S.~government sources.  However, the U.S.~Department of Defense had not confirmed the event, as was required by TradeSports' contract.  TradeSports concluded that the contract's conditions had not been met, and paid out accordingly.\footnote{See \texttt{https://en.wikipedia.org/wiki/Intrade\#Disputes} for details.}

This is a case where the spirit of the market -- to predict the missile launch -- was clearly satisfied, but the letter of the market -- to predict whether the U.S.~Department of Defense would confirm the launch -- was not.  TradeSports, being a centralized website, was able to unilaterally declare the outcome of the market.  If such a situation arises in an Augur market, REP holders may have differing opinions about how the market should resolve, and stake their REP accordingly.  In the worst case, this could result in a fork where REP in more than one child universe maintains a non-zero market value.

\begin{acknowledgments}\label{section:acknowledgements}
We thank Abraham Othman, Alex Chapman, Serena Randolph, Tom Haile, George Hotz, Scott Bigelow, and Peronet Despeignes for their helpful feedback and suggestions.
\end{acknowledgments}

\cleardoublepage

\bibliographystyle{unsrt}
\bibliography{augur}

\begin{appendix}

\cleardoublepage

\section{Finalization Time \& Redistribution}\label{section:finalization_time}

We begin with some notation, definitions, and observations.

\begin{definition}
For a given market $M$, let ${\Omega}_M$ be the outcome space (or set of outcomes) of $M$.
\end{definition}

\begin{definition}
For $n \geq 1$ and $\omega \in {\Omega}_M$,  let $S(\omega,n)$ denote the total amount of stake on outcome $\omega$ at the beginning of dispute round $n$. This includes all stake from all successful dispute bonds in favor of $\omega$ over all previous dispute rounds.
\end{definition}

\begin{definition}
For $n \geq 1$ and $\omega \in {\Omega}_M$, let $S(\overline{\omega},n)$ denote the amount of stake on all outcomes in ${\Omega}_M$ \emph{except for} $\omega$ at the beginning of dispute round $n$:
\[ S(\overline{\omega},n)= \sum_{\overset{\gamma \in {\Omega}_M}{\gamma \neq \omega}}^{} S(\gamma,n) \]
\end{definition}

\begin{definition}
For $n \ge 1$, let $A_n$ denote the total stake over all outcomes $M$ at the beginning of dispute round $n$: \[ A_n = \sum_{\omega \in {\Omega}_M}^{} S(\omega,n) \]
\end{definition}

\begin{observation}\label{ob:stake_partition}
It follows that $A_n - S(\omega,n) = S(\overline{\omega},n)$.
\end{observation}

\begin{definition}
For $n \geq 1$, let $\hat{\omega}_{n}$ denote the tentative outcome at the beginning of dispute round $n$.  For example, $\hat{\omega}_{1}$ is the outcome reported by the initial reporter.
\end{definition}

\begin{definition}
For $n \geq 1$ and $\omega \neq \hat{\omega}_{n}$, let $B(\omega,n)$ denote the amount of stake required to successfully fill a dispute bond in favor of outcome $\omega$ during dispute round $n$.
\end{definition}

Recall that the amount of stake required to successfully fill a dispute bond in favor of outcome $\omega$ during dispute round $n$, where $\omega \neq \hat{\omega}_{n}$ is given by Eq.~\ref{eq:bond_size}, $B(\omega,n) = 2A_{n} - 3S(\omega,n)$.

\begin{observation}\label{ob:only_successfull_bond_stake_applies_v1}
If a dispute bond is successfully filled in favor of outcome $\omega$ during dispute round $n$, then $S(\omega,n+1)=B(\omega,n)+S(\omega,n)$.  That is, the successful dispute stake is the only new stake applied to outcome $\omega$ at the end of dispute round $n$.
\end{observation}

\begin{observation}\label{ob:only_successfull_bond_stake_applies_v2}
For all $\omega \neq \hat{\omega}_{n}, S(\omega,n-1)=S(\omega,n)$.  That is, if a dispute bond is not entirely filled in favor of outcome $\omega$, then no additional stake is added to outcome $\omega$ at the beginning of the next dispute round.  This is due to the fact that all unsuccessful dispute stake is returned to the users at the end of the dispute round.
\end{observation}

\begin{observation}\label{ob:only_successfull_bond_stake_applies_v3}
For all $n \geq 2, A_{n} = A_{n-1} + B(\hat{\omega}_{n},n-1)$.  That is, the total stake over all outcomes at the beginning of a dispute round is simply the total stake from the beginning of the previous dispute round plus the successful dispute stake from the previous dispute round.  All other stake is returned to users at the end of the previous dispute round.
\end{observation}

\begin{lemma}\label{le:tentative_outcomes_are_a_third_of_all_stake}
$S(\hat{\omega}_{n},n) = 2S(\overline{\hat{\omega}_{n}},n)$, for $n \geq 2$.
\end{lemma}

\begin{proof}
Suppose a market enters dispute round $n$, where $n \geq 2$.  During dispute round $n-1$, the outcome $\hat{\omega}_{n-1}$ must have been successfully disputed in favor of outcome $\hat{\omega}_{n}$.  According to Eq.~\ref{eq:bond_size}, the size of that dispute bond is $B(\hat{\omega}_{n},n-1) = 2A_{n-1} - 3S(\hat{\omega}_{n},n-1)$.  Using observation \ref{ob:stake_partition}, this can be rewritten as
\beq \label{eq:bond_size_stake_partition}
B(\hat{\omega}_{n},n-1) + S(\hat{\omega}_{n},n-1) = 2S(\overline{\hat{\omega}_{n}},n-1)
\eeq

We know the dispute bond was successfully filled during round $n-1$.  Using observation \ref{ob:only_successfull_bond_stake_applies_v1}, we see that $B(\hat{\omega}_{n},n-1) + S(\hat{\omega}_{n},n-1) = S(\hat{\omega}_{n},n)$.  Observation \ref{ob:only_successfull_bond_stake_applies_v2} tells us that the total amount staked on $\overline{\hat{\omega}_{n}}$ is unchanged from round $n-1$ to $n$, $2S(\overline{\hat{\omega}_{n}},n-1) = 2S(\overline{\hat{\omega}_{n}},n)$.  Thus, Eq.~\ref{eq:bond_size_stake_partition} reduces to $S(\hat{\omega}_{n},n) = 2S(\overline{\hat{\omega}_{n}},n)$.
\end{proof}

\begin{theorem}\label{th:roi_guarantee}
Any REP holders successfully disputing an outcome in favor of a market's final outcome will receive a 40\% ROI on their dispute stake (measured in REP that exists in a universe that corresponds to the market's final outcome), unless the market is interrupted by some other market causing a fork.
\end{theorem}

\begin{proof}
During a fork, all users who successfully filled dispute bonds in favor of the forking market's final outcome are given (via coins minted during the fork) a 40\% return on their dispute stake when they migrate their dispute stake to the corresponding child universe. Thus, in the case where the market in question has caused a fork, the theorem is immediately true.

Now consider the case where the market in question resolves without causing a fork, and reporting is not interrupted by some other market causing a fork.

Denote the market's final outcome by $\omega_{\mathrm{Final}}$ and suppose the market resolves at the end of dispute round $n$, where $n \geq 2$. That means the tentative outcome for round $n$ is $\omega_{\mathrm{Final}}$, and that outcome is not successfully disputed during round $n$. In other words: $\hat{\omega}_{n} = \omega_{\mathrm{Final}}$. Then by Lemma \ref{le:tentative_outcomes_are_a_third_of_all_stake} we know that
$ S(\omega_{\mathrm{Final}},n) = 2S(\overline{\omega_{\mathrm{Final}}},n)$.

Since the market resolves at the end of round $n$ with no further stake added to any outcome, the above equation shows the final amount of stake on the market's final outcome, $\omega_{\mathrm{Final}}$, and the sum of all stake on all of the market's other outcomes, $\overline{\omega_{\mathrm{Final}}}$. Note that there is exactly twice as much stake on the market's final outcome as there is on all other outcomes combined.

Augur burns 20\% of the all stake on the non-final outcomes and redistributes the rest to users who staked on $\omega_{\mathrm{Final}}$, in proportion to the amount of REP they staked. Therefore the users who successfully filled a dispute bond in favor of $\omega_{\mathrm{Final}}$ get a 40\% ROI on their staked REP.
\end{proof}

Next, consider the maximum number of dispute rounds required to resolve a market.  Eq.~\ref{eq:bond_size} is minimized when $\omega$ is chosen to be the non-tentative outcome that begins the dispute round with the greatest amount of stake.  Lemma \ref{le:tentative_outcomes_are_a_third_of_all_stake} implies that the non-tentative outcome with the greatest amount of stake is the previous dispute round's tentative outcome.  Therefore, the smallest possible dispute bond size that can be successfully filled during dispute round $n$, where $n \geq 2$, is $B(\hat{\omega}_{n-1},n)$.

In other words, the dispute bond size grows \emph{slowest} when the same two outcomes are repeatedly disputed in favor of one another.  It follows that the number of dispute rounds required for a market to initiate a fork is \emph{maximized} when the same two outcomes are repeatedly disputed in favor of one another.  Therefore we can determine the maximum number of dispute rounds that any market may undergo before initiating a fork by finding the maximum number of dispute rounds that can occur in the particular case where the same two market outcomes are repeatedly disputed in favor of one another.  We examine that case now.

Suppose that every successful dispute bond is filled in favor of the previous dispute round's tentative outcome.  Then the two tentative outcomes that are iteratively disputed in favor of one another other are $\hat{\omega}_{1}$ and $\hat{\omega}_{2}$.

\begin{observation}\label{ob:tentative_outcomes_repeat}
In the case where the same two tentative outcomes are repeatedly disputed in favor of one another, $\hat{\omega}_{n} = \hat{\omega}_{n-2}$ for all $n \geq 3$.
\end{observation}

\begin{definition}
Let $d$ denote the amount of stake placed on $\hat{\omega}_{1}$ during the initial report.  Because the tentative outcome for each round is known in this situation, we can simplify our notation for the dispute bond sizes.  Define a shorthand $B_n$ to denote the bond size required for round $n$, so that $B_{1} = 2d$ and $B_{n} = B(\hat{\omega}_{n-1},n)$ for all $n \geq 2$.  This will make for easier reading and comprehension.
\end{definition}

\begin{observation}\label{ob:every_other_dispute_stake_applies_to_same_outcome}
In the case where the same two tentative outcomes are repeatedly disputed in favor of one another, $S(\hat{\omega}_{n-1},n) = S(\hat{\omega}_{n-1},n-2) + B_{n-2}$ for $n \geq 3$.  (That is, every other successful dispute bond is added to the same outcome.)
\end{observation}

\begin{lemma}\label{le:bond_sizes}
If the same two tentative outcomes are repeatedly disputed in favor of one another, then for all $n$ where $n \geq 3$:
\begin{enumerate}
\item{$S(\hat{\omega}_{n-1},n)=\frac{2}{3}B_{n-1}$}
\item{$A_{n} = 2B_{n-1}$ and}
\item{$B_{n} = 3d2^{n-2}$}
\end{enumerate}
\end{lemma}

\begin{proof}
(By induction on $n$)

Suppose the same two tentative outcomes are repeatedly disputed in favor of one another.

(Base Case)
By definition and Eq.~\ref{eq:bond_size} we make the following observations.

\begin{itemize}
\item{$S(\hat{\omega}_{1},1)=d$, $S(\hat{\omega}_{2},1)=0$, $A_{1}=d$, and $B_{1}=2d$}
\item{$S(\hat{\omega}_{1},2)=d$, $S(\hat{\omega}_{2},2)=2d$, $A_{2}=3d$, and $B_{2}=3d$}
\item{$S(\hat{\omega}_{1},3)=4d$, $S(\hat{\omega}_{2},3)=2d$, $A_{3}=6d$, and $B_{3}=6d$}
\end{itemize}

$S(\hat{\omega}_{3-1},3)=S(\hat{\omega}_{2},3)=2d=\frac{2}{3}(3d)=\frac{2}{3}B_{2}=\frac{2}{3}B_{3-1}$, so part 1 of the lemma holds for $n=3$.

$A_{3}=6d=2(3d)=2B_{2}=2B_{3-1}$, so part 2 of the lemma holds for $n=3$.

$B_{3}=6d=3d2^{3-2}$, so part 3 of the lemma holds for $n=3$.

\vspace{4mm}

Therefore the lemma, in its entirety, holds true for the base case of $n=3$.

\vspace{4mm}

(Induction)
Suppose the lemma is true for all $n$ such that $3 \leq n \leq k$.  We want to show that the lemma holds for $n=k+1$.  That is, we want to show that:

\begin{enumerate}[label=(\alph*)]
\item{$S(\hat{\omega}_{k},k+1)=\frac{2}{3}B_{k}$}
\item{$A_{k+1}=2B_{k}$ and}
\item{$B_{k+1}=3d2^{k-1}$}
\end{enumerate}

First, we prove part (a). By observation \ref{ob:every_other_dispute_stake_applies_to_same_outcome}:
\[ S(\hat{\omega}_{k},k+1)=S(\hat{\omega}_{k},k-1)+B_{k-1} \]

By observation \ref{ob:tentative_outcomes_repeat} we can rewrite the above as:
\[ S(\hat{\omega}_{k-2},k+1)=S(\hat{\omega}_{k-2},k-1)+B_{k-1} \]

By the induction hypothesis, we can rewrite $S(\hat{\omega}_{k-2},k-1)$ as $\frac{2}{3}B_{k-2}$ on the right-hand side to get:
\[ S(\hat{\omega}_{k-2},k+1)=\tfrac{2}{3}B_{k-2}+B_{k-1} \]

By the induction hypothesis, we can write $B_{k-2}$ as $3d2^{k-4}$ and $B_{k-1}$ as $3d2^{k-3}$:
\[ S(\hat{\omega}_{k-2},k+1)=d2^{k-1} \]

Applying observation \ref{ob:tentative_outcomes_repeat} to the left-hand side we get:
\[ S(\hat{\omega}_{k},k+1)=d2^{k-1} \]

Finally, note that by the above equation and the induction hypothesis, $S(\hat{\omega}_{k},k+1)=d2^{k-1}=\frac{2}{3}(3d2^{k-2})=\frac{2}{3}B_{k}$. This proves part (a).

Next, we prove part (b). By observation \ref{ob:only_successfull_bond_stake_applies_v3}:
\[ A_{k+1}=A_{k}+B_{k} \]

By the induction hypothesis, $A_{k}=2B_{k-1}$:
\[ A_{k+1}=2B_{k-1}+B_{k} \]

By the induction hypothesis, $B_{k-1}=3d2^{k-3}$, so the right-hand side can be simplified to
\[ A_{k+1}=3d2^{k-2}+B_k \]

By the induction hypothesis, $B_{k}=3d2^{k-2}$ to rewrite the right-hand side as
\[ A_{k+1}=2B_{k}, \]
and part (b) is proved.

Finally, we prove part (c).  By Eq.~\ref{eq:bond_size}:
\[ B_{k+1}=2A_{k+1}-3S(\hat{\omega}_{k},k+1) \]

By observation \ref{ob:every_other_dispute_stake_applies_to_same_outcome}, we can write $S(\hat{\omega}_{k},k+1)$ as $S(\hat{\omega}_{k},k-1)+B_{k-1}$:
\[ B_{k+1}=2A_{k+1}-3\left(S(\hat{\omega}_{k},k-1)+B_{k-1}\right) \]

By observation \ref{ob:tentative_outcomes_repeat}, $\hat{\omega}_{k}=\hat{\omega}_{k-2}$:
\[ B_{k+1}=2A_{k+1}-3\left(S(\hat{\omega}_{k-2},k-1)+B_{k-1}\right) \]

By observation \ref{ob:only_successfull_bond_stake_applies_v3}, $A_{k+1}=A_{k}+B_{k}$:
\[ B_{k+1}=2\left(A_{k}+B_{k}\right)-3\left(S(\hat{\omega}_{k-2},k-1)+B_{k-1}\right) \]

By the induction hypothesis, $A_{k}=2B_{k-1}$ and $S(\hat{\omega}_{k-2},k-1)=\frac{2}{3}B_{k-2}$:
\[ B_{k+1}=2\left(2B_{k-1}+B_{k}\right)-3\left(\tfrac{2}{3}B_{k-2}+B_{k-1}\right) \]

By the induction hypothesis, $B_{k}=3d2^{k-2}$, $B_{k-1}=3d2^{k-3}$ and $B_{k-2}=3d2^{k-4}$.  Making these substitutions and simplifying yields:
\[ B_{k+1}=3d2^{k-1} \]

This proves part (c), and concludes the proof of the lemma.
\end{proof}

\begin{theorem}\label{th:twenty_rounds}
If not interrupted by some other market causing a fork, a given market may undergo at most 20 dispute rounds before finalizing or causing a fork.
\end{theorem}

\begin{proof}
Suppose that a given market is not interrupted by some other market causing a fork. Then, as shown above, we know that the number of dispute rounds required for a market to initiate a fork is maximized when the same two outcomes are repeatedly disputed in favor of one another.  Part 3 of Lemma \ref{le:bond_sizes} tells us that, in this situation, the dispute bond size required for successfully disputing the tentative outcome during round $n$ is given by $3d2^{n-2}$, where $d$ is the amount of the stake placed during the initial report.

We know that forks are initiated after the successful fulfillment of a dispute bond with size at least 2.5\% of all existing REP, and we know that there are 11 million REP in existence. Thus a fork is initiated when a dispute bond of size 275,000 REP is filled. We also know that $d \geq 0.35$ REP, because the minimum amount of stake on the initial report is $0.35$ REP\footnote{See appendixes \ref{section:bond_size_adjustment_details_no-show_bonds} and \ref{section:bond_size_adjustment_details_designated_reporter_stake}}.

Solving $3(0.35)2^{n-2}>275,000$ for $n \in \mathbb{Z}$ yields $n \geq 20$.  Thus, we can guarantee that a market will resolve or cause a fork after at most 20 dispute rounds.
\end{proof}

\section{Bond Size Adjustments}\label{section:bond_size_adjustment_details}

The validity bond and the creation bond are dynamically adjusted based on the behavior of participants during the previous dispute window. The creation bond is the maximum of the \textit{no-show bond} and the \textit{designated reporter bond} -- two values that the system tracks in order to compute the creation bond but does not expose to users.

Here we describe how we adjust these bonds.

We define the function $f: [0, 1] \rightarrow [\frac{1}{2}, 2]$ by:\footnote{This formula may change once empirical data from live markets is obtained.}
\beq\label{eq:validity_bond_adjustment}
f(x) = \begin{cases} 
    \frac{100}{99} x + \frac{98}{99} & \text{for}\quad x > \frac{1}{100} \\
    50x + \frac{1}{2} & \text{for}\quad x \leq \frac{1}{100}
\end{cases}
\eeq

The function $f$ is used to determine the multiple used in these adjustments, as described in the subsections below. In brief, if the undesirable behavior occurred exactly 1\% of the time during the previous dispute window, then the bond size remains the same. If it was less frequent, then the bond size will be reduced by as much as half. If it was more frequent, then the bond size will be increased by as much as a factor of 2.

\subsection{Validity Bond}\label{section:bond_size_adjustment_details_validity_bonds}

During the very first dispute window after launch, the validity bond will be set at 0.01 ETH. Then, if more than 1\% of the finalized markets in the previous dispute window were invalid, the validity bond will be increased.  If less than 1\% of the finalized markets in the previous dispute window were invalid, then the validity bond will be decreased (but will never be lower than 0.01 ETH).

In particular, we let $\nu$ be the proportion of finalized markets in the previous dispute window that were invalid, and $b_v$ be the amount of the validity bond from the previous dispute window. Then the validity bond for the current window is $\max\left\{\frac{1}{100}, b_v f(\nu)\right\}$.

\subsection{No-Show Bond}\label{section:bond_size_adjustment_details_no-show_bonds}

During the very first dispute window after launch, the no-show bond will be set at 0.35 REP. As with the validity bond, the no-show bond is adjusted up or down, targeting a 1\% no-show rate with a floor of 0.35 REP.

Specifically, we let $\rho$ be the proportion of markets in the previous dispute window whose designated reporters failed to report on time, and we let $b_r$ be amount of the no-show bond from the previous dispute window. The the amount of the no-show bond for the current dispute window is $\max\left\{0.35,b_r f(\rho)\right\}$.

\subsection{Designated Reporter Bond}\label{section:bond_size_adjustment_details_designated_reporter_stake}

During the very first dispute window after launch, the amount of the designated reporter bond will be set at 0.35 REP. The amount of the designated reporter bond is dynamically adjusted according to how many designated reports were incorrect (failed to concur with the final market outcome) during the previous dispute window.

In particular, we let $\delta$ be the proportion of designated reports that were incorrect during the previous dispute window, and we let $b_d$ be the amount of the designated reporter stake during the previous dispute window, then the amount of the designated reporter bond for the current window is $\max\left\{0.35, b_d f(\delta)\right\}$.

\end{appendix}

\end{document}